\documentclass[journal=jpclcd,manuscript=letter]{achemso}
\usepackage{amsmath,amssymb}
\usepackage{float}
 \usepackage{tabularx}
 \usepackage{arydshln}
\usepackage{upgreek}
\usepackage{adjustbox}
\usepackage{color}
\usepackage{array}
\usepackage[version=3]{mhchem} 

\author{Pooja Basera}
\email{Pooja.Basera@physics.iitd.ac.in[PB]}
\author{Saswata Bhattacharya}
\email{saswata@physics.iitd.ac.in[SB]}
\phone{+91-11-2659 1359}
\affiliation[Indian Institute of Technology Delhi]
{Department of Physics, Indian Institute of Technology Delhi, New Delhi, India}
\title[An \textsf{achemso} demo]
{Chalcogenide perovskites: an emerging class of semiconductors for optoelectronics}
\begin{document}
\begin{abstract}
Chalcogenide perovskites have received considerable interest in photovoltaic research community owing to their stability (thermal and aqueous), non-toxicity and lead free composition. However, to date 
a theoretical study mainly focusing on the excitonic and polaronic properties are not explored rigorously, due to its huge computational demand. Herein, we capture the excitonic and polaronic effects in a series of chalcogenide perovskites ABS$_3$ where A=Ba, Ca, Sr, and B=Hf, Sn by employing state-of-the art hybrid density functional theory and many body perturbative approaches viz. GW and BSE. We find that these perovskites possess a large exciton binding energy than 3D inorganic-organic hybrid halide perovskites. We examine the interplay of electronic and ionic contribution to the dielectric screening, and conclude that electronic contribution is dominant over the ionic contribution. Further using Feynman polaron model, polaron parameters are computed, and we observe that charge separated polaronic states are less stable than bound excitons. Finally, the theoretically calculated spectroscopic limited maximum efficiency (SLME) suggests that among all chalcogenide perovskites, CaSnS$_3$ could serve as a best choice for photovoltaic applications.

  \begin{tocentry}
  \begin{figure}[H]%
  \includegraphics[width=1.0\textwidth]{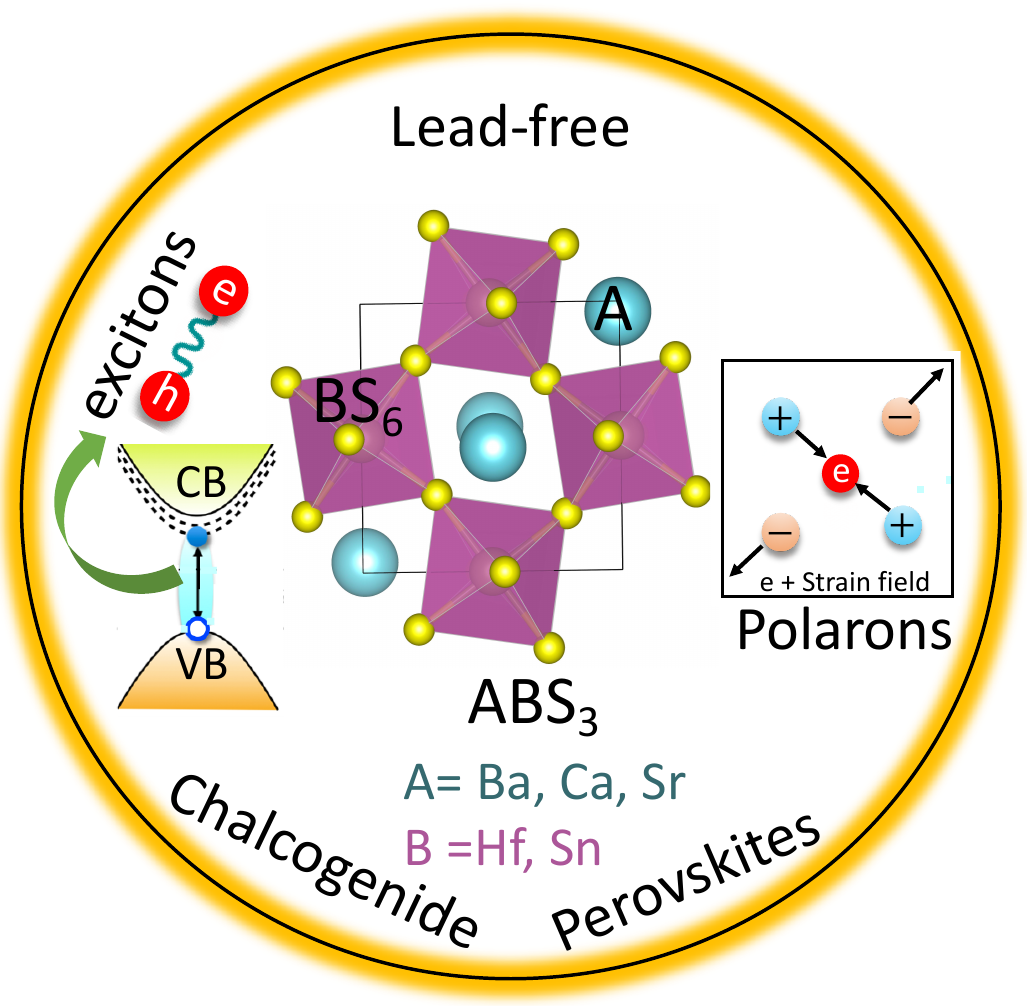}
 \end{figure}	
 \end{tocentry}
\end{abstract}
Perovskites have gained unprecedented attention in the past few years owing to their versatile chemical and physical properties, and lead to a breakthrough progress from 3.8\% to 25.5\% ~\cite{doi:10.1021/ja809598r,nrel}. This indicates that device performance is intimately linked with basic material properties. The suitable material characteristics allow to significantly shorten the time period for the evolution of high-efficiency cells. However, the toxicity, degradability and the instability are still serious challenges for large scale industrial applications~\cite{babayigit2016toxicity,pooja-prb}. This motivates researchers towards a new species of perovskites, namely, chalcogenide perovskites that possess Pb-free composition, are earth-abundant, and highly stable in ambient atmosphere~\cite{kuhar2017sulfide,tiwari2021chalcogenide}. Several studies have been performed experimentally as well as theoretically that show the successful formation of chalcogenide perovskites along with their promising properties~\cite{https://doi.org/10.1107/S056774088000845X,LEE20051049,doi:10.1021/acs.chemmater.5b04213,PERERA2016129,https://doi.org/10.1002/adma.201604733,doi:10.1021/acsaem.9b02428,doi:10.1021/acsomega.0c00740,D0NR08078K,doi:10.1021/acs.chemmater.5b04213,C7EE02702H,C9TA03116B,ma13040978,PhysRevMaterials.3.101601}. Perera $et$ $al.$ have synthesised AZrS$_3$ (A = Ba, Ca, Sr) using high temperature sulfurization of the oxides with CS$_2$~\cite{perera2016chalcogenide}. S. Niu $et$ $al.$ have also synthesized them as well as probed the optoelectronic properties experimentally~\cite{niu2017bandgap}. Meng $et$ $al.$ predicted mixed perovskites chalcogenides BaZr$_{1-x}$Ti$_x$S$_3$ by repeated  annealing of binary mixtures~\cite{meng2016alloying}. Sun $et$ $al.$ proposed S and Se based perovskites and predicted their suitable band gaps for single junction solar cell~\cite{sun2015chalcogenide}. However, most of the earlier experimental and theoretical studies are mainly focused on Zr-based chalcogenide perovskites~\cite{kumar2021optoelectronic}. Recently, a  few reports in the literature presented a first-principles studies for Sn based chalcogenides as a high performance thermoelectric materials~\cite{li2022weak}, and Hf based chalcogenides as green light emitting semiconductors~\cite{hanzawa2019material}. In addition, a recent experimental studies successfully unveil the synthesis of CaSnS$_3$ perovskite~\cite{shaili2021synthesis}. In view of this, we aim to provide a detailed theoretical study for Hf and Sn- based chalcogenides concerning the properties suitable for optoelectronic applications that may act as valuable guidance for the ongoing experimental work. \\
Intriguingly, the charge separation in these materials are considerably affected by the exciton formation. The operation behind the solar cell mechanism depends on the thermally dissociate excitons into free electrons and holes, which in turn induce free-charge transport. Therefore, the accurate determination of the exciton binding energy is crucial for their active usage in optoelectronic materials. Additionally, the charge carrier and exciton dynamics are also influenced by the polaron formation. The multiple photophysical phenomena in these materials can be explained by electron-phonon interactions and the transport of polarons~\cite{miyata2018ferroelectric,ponce2019origin}. Regardless of several studies on chalcogenide perovskites, a thorough understanding on excitonic and polaronic effects on Hf and Sn-based chalcogenides are hithertho unknown, and unveil for the first time, to the best of our knowledge.\\
In this Letter, we have performed a robust study of electronic, optical, excitonic and polaronic properties of chalcogenide perovskites ABS$_3$ (where A= Ba, Ca, Sr, and B=Hf, Sn) within the framework of density functional theory (DFT)~\cite{PhysRev.136.B864,PhysRev.140.A1133}, hybrid DFT (HSE06) and beyond DFT approaches (GW and BSE). First, we optimize the crystal structures using semi-local PBE~\cite{PhysRevLett.77.3865} exchange-correlation ($\epsilon_\textrm{xc}$) functional, and validate the optimized lattice parameters in light of the available experimental results. Then, to study electronic structure, atom-projected electronic partial density of states (pDOS) are computed using hybrid $\epsilon_\textrm{xc}$ functional HSE06~\cite{doi:10.1063/1.2404663}. Subsequently, the optical properties are determined using many-body perturbation theory (MBPT). Then, we solve the Bethe-Salpeter equation (BSE)~\cite{PhysRevLett.80.4510,PhysRevLett.81.2312} on top of single-shot G$_0$W$_0$@PBE~\cite{PhysRev.139.A796,PhysRevLett.55.1418}, to obtain the exciton binding energy as well as electronic contribution to the dielectric function. On the other hand, the ionic contribution to the dielectric function is determined using density functional perturbation theory (DFPT). Finally, we compute the spectroscopic limited maximum efficiency (SLME)~\cite{PhysRevLett.108.068701} using the quasiparticle (QP) band gap and absorption coefficient.

\begin{figure}[H]
	\centering
	\includegraphics[width=1.0\textwidth]{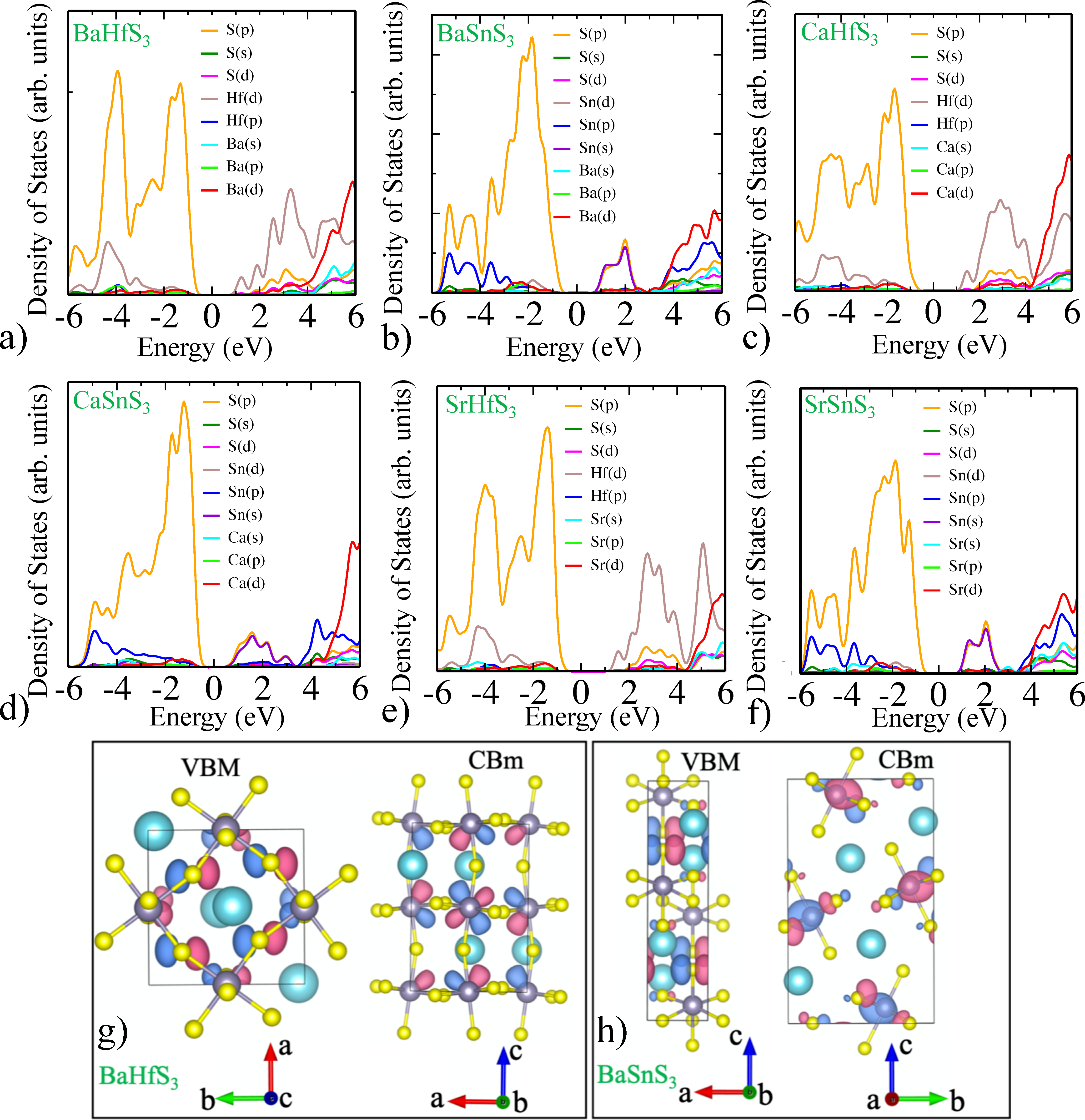}
	\caption{Electronic partial density of states (pDOS) of (a) BaHfS$_3$, (b) BaSnS$_3$, (c) CaHfS$_3$, (d) CaSnS$_3$, (e) SrHfS$_3$, (f) SrSnS$_3$ using HSE06 $\epsilon_\textrm{xc}$ functional. Wave function analysis in real space of (g) BaHfS$_3$, and (h) BaSnS$_3$ at VBM and CBm.}
	\label{fig_dos}
\end{figure}

Herein, we consider the distorted orthorhombic phase of chalcogenide perovskites ABS$_3$ (A=Ba, Ca, Sr, and B=Hf, Sn) having the space group $Pnma$~\cite{https://doi.org/10.1107/S056774088000845X} (see Figure \ref{fig_dos}a). For BaSnS$_3$ and SrSnS$_3$, the needle-like phase is examined in our study. Table \ref{tbl:1} presents the PBE $\epsilon_\textrm{xc}$ functional optimized lattice parameters, which are closely in agreement with experimental results reported in literature.
\begin{figure}[h]
	\centering
	\includegraphics[width=1.0\textwidth]{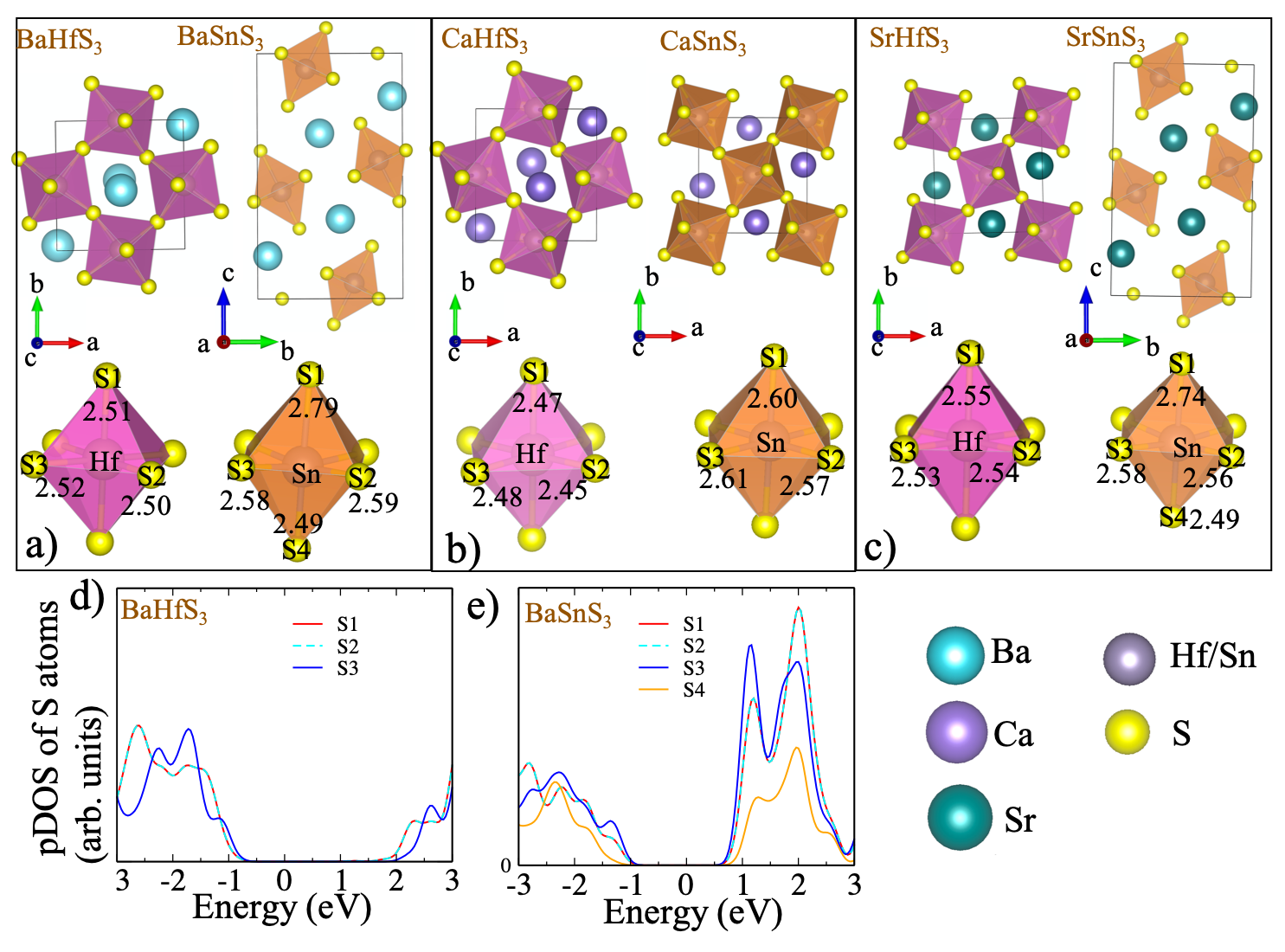}
	\caption{Crystal structure (a) BaHfS$_3$ in orthorhombic distorted phase and BaSnS$_3$ in needle-like phase. (b) CaHfS$_3$ and CaSnS$_3$ in orthorhombic distorted phase. (c) SrHfS$_3$ in orthorhombic distorted phase and SrSnS$_3$ in needle-like phase. Sulphur atom projected density of states (d) BaHfS$_3$ and (e) BaSnS$_3$}
	\label{structure}
\end{figure}
\begin{table}
	\caption{Calculated lattice parameters of ABS$_3$ (A=Ba, Ca, Sr and B = Hf, Sn) chalcogenide perovskites. The experimental values are provided in brackets, and are taken from Ref. ~\cite{https://doi.org/10.1107/S056774088000845X,yamaoka1970preparations,shaili2021synthesis}. Calculated distortion parameters for BS$_6$ octahedra in ABS$_3$ chalcogenide perovskites.}
\renewcommand{\arraystretch}{0.9}
\resizebox{\textwidth}{!}{
	\label{tbl:1}
	\centering
	\setlength\extrarowheight{+3pt}
	\begin{tabular}[c]{ccccc}
		\hline
		Configurations & a (\AA) & b (\AA)&  c (\AA) \\
		\hline
		BaHfS$_3$ & 6.95 (6.99)& 6.99 (7.00)& 9.99 (9.92)\\
		BaSnS$_3$ &3.97 (3.93)& 8.48 (8.53)&15.35 (14.51)\\
		CaHfS$_3$ &6.39 (6.52)&6.89 (6.98)&9.35 (9.54)\\
		CaSnS$_3$ &6.72 (6.68)&7.06(7.08)&9.67 (11.28)\\
		SrHfS$_3$ &6.77 (6.72)&7.09 (7.05)&9.77 (9.72)\\
		SrSnS$_3$ &3.78 (3.79)&8.41 (8.74)& 14.62(14.08)\\
		\hline
Octahedra (BS$_6$) & Average bond length (\AA)  &  Bond angle variance (deg.($^2$)) & Polyhedral volume (\AA$^3$) & Distortion index \\
		\hline
		BaHfS$_3$ (HfS$_6$) &2.51& 0.29& 21.04 & 0.0036\\
		BaSnS$_3$ (SnS$_6$) &2.61 & 19.61&23.38 &0.024\\
		CaHfS$_3$ (HfS$_6$)&2.47&1.14&20.06&0.00513\\
		CaSnS$_3$(SnS$_6$) &2.59&3.22&23.23&0.00513\\
		SrHfS$_3$ (HfS$_6$)&2.54 &0.15&21.86&0.00377\\
		SrSnS$_3$ (SnS$_6$) &2.59&15.72& 22.92&0.01958\\
		\hline
	\end{tabular}}
\end{table}
In addition, the octahedra distortions (BS$_6$) for ABS$_3$ chalcogenides are also listed in Table\ref{tbl:1}. The average bond length, bond angle variance, polyhedral volume and distortion index are calculated corresponding to BS$_6$ octahedra. The data shows that the BS$_6$ octahedra in chalcogenides having needle like phase (BaSnS$_3$ and SrSnS$_3$) is more distorted compared to octahedra in orthorhombic phase (see Fig~\ref{structure} and Table \ref{tbl:1}). In BaSnS$_3$ and SrSnS$_3$, there are four types of Sn-S bonds of different strengths, namely S1, S2, S3, S4 (see Fig~\ref{structure}[a-c]). We notice that Sn-S1 bonding is relatively weaker, however Sn-S4 bonding is the strongest one. The atom-projected density of states are calculated to disentangle the contribution of each S atoms at valence band edges (Note that S atom has main contribution at valence band maximum for these perovskites (vide infra)). For BaHfS$_3$, the S1, S2 and S3 contribution are same at valence band edges, because of almost same binding strength with Hf atom, however for BaSnS$_3$, Sn-S4 has stronger binding that leads to its least contribution at valence band edges (see SI, Fig S3, for remaining configurations). Therefore, the differentiation of S atoms at valence band edges subsequently lead to lighter valence bands (low m$_\textrm{h}^*$), which eventually helps to boost the hole mobility (Table \ref{tbl:2}). \\
To gain deep insights on the electronic structure, we compute the electronic pDOS of the ABS$_6$ chalcogenides using HSE06 $\epsilon_\textrm{xc}$ functional as shown in Figure \ref{fig_dos}[a-f]. For AHfS$_3$ perovskites, the valence band maximum (VBM) is majorly contributed by S 3p-orbitals, whereas Hf 4d-orbitals primarily contributed at the conduction band minimum (CBm). The remaining orbitals have minor contribution at VBM and CBm. For ASnS$_3$ perovskites, the VBM is mainly contributed by S 3p -orbitals, the CBm is dominated by the hybridization of S (3p) and Sn (5s) orbitals. The presence of hybridized states in case of ASnS$_3$ perovskite is responsible for its lower band gap than AHfS$_3$ perovskites, the latter does not contain hybridized states. The orbitals and their hybridzation can be visualized clearly from wave function analysis. For BaHfS$_3$, the S p orbitals and the Hf d orbitals contribution at VBM and CBm, respectively are shown in Figure \ref{fig_dos}[g]. For BaSnS$_3$, the contribution of S p orbitals at VBM and the hybridization of Sn s and S p orbitals are shown in Figure \ref{fig_dos}[h]. From the electronic structure, we observe p-d transition in Hf-based perovskites, where as p-p and p-s  transition in Sn-based perovskites. Notably, perovskites are well known for their transport properties. Thus, here we compute the band structures for all the cases (see SI, Fig S1) and their corresponding reduced mass ($\mu$). We observe $\mu$ $<$ 1 for all the cases, which in turn indicates high carrier mobility, and therefore infers the better charge carrier transport. However, for BaSnS$_3$ and SrSnS$_3$, m$_\textrm{h}^*$ $<$ m$_\textrm{e}^*$  due to the difference in the Sn-S bonding strength at valence band edge (vide supra).

\begin{table}
	\caption{Calculated effective mass of electron (m$_\textrm{e}^*$), hole (m$_\textrm{h}^*$) and reduced mass ($\mu$) along a $\Gamma-$Z high symmetry path. All  values are in terms of free-electron mass (m$_\textrm{e}$).}
	\label{tbl:2}
	\centering
	\setlength\extrarowheight{+4pt}
	\begin{tabular}[c]{cccc}
		\hline
		Configurations & m$_\textrm{e}^*$ & m$_\textrm{h}^*$ &  $\mu$\\
		\hline
		BaHfS$_3$ & 0.367& 0.642& 0.233\\
		BaSnS$_3$ &0.639& 0.233&0.170\\
		CaHfS$_3$ &0.444&0.500&0.235\\
		CaSnS$_3$ &0.385&0.450&0.207\\
		SrHfS$_3$ &0.445&0.567&0.249\\
		SrSnS$_3$ &0.527&0.213&0.152\\
		\hline
	\end{tabular}
\end{table}
\begin{table}
	\caption{Band gap (in eV) of chalcogenide perovskites. $i$, $d$, $e$ and $t$ represent indirect, direct, experimental and theoretical band gap.}
	\label{tbl:3}
	\centering
	\setlength\extrarowheight{+4pt}
	\begin{tabular}{ccccc}
		\hline
		Configurations &PBE& HSE06 & G$_0$W$_0$@PBE &  Previous work\\
		\hline
		BaHfS$_3$ &1.12& 1.97& 2.17& 2.17$_e$\cite{PERERA2016129}\\
		BaSnS$_3$ &0.87$^i$(0.95$^d$) & 1.83$^i$(1.96$^d$)&1.91$^i$(2.03$^d$)&1.62$_t$\cite{https://doi.org/10.1002/adma.201604733}\\
		CaHfS$_3$&1.38 &2.23&2.46&2.21$_t$\cite{https://doi.org/10.1002/adma.201604733}\\
		CaSnS$_3$ &0.76 &1.40&1.43&1.58$_t$\cite{https://doi.org/10.1002/adma.201604733}\\
		SrHfS$_3$ &1.47 &2.32&2.59&2.41$_e$\cite{https://doi.org/10.1002/adma.201604733}\\
		SrSnS$_3$ &0.96 &1.94&2.04&1.56$_t$\cite{https://doi.org/10.1002/adma.201604733}\\
		\hline
	\end{tabular}
\end{table}
Note that, all the ABS$_3$ perovskites have band gap less than 2.6 eV, thereby suitable for optoelectronic applications. However, we observe a slight indirect band gap for BaSnS$_3$ (differ by 0.08 eV), and weakly indirect band gap for SrSnS$_3$ (differ by few meV's). In other words, the direct gap is at energy level very close to indirect gap. The reason for indirect band gap could be a large octahedral tilting (BS$_6$) for needle like phase, indicated from distortion index value given in Table \ref{tbl:1}. The band gaps computed from semi-local PBE $\epsilon_\textrm{xc}$ functional are underestimated due to well known self-interaction error (see Table \ref{tbl:3}). However, hybrid $\epsilon_\textrm{xc}$ functional HSE06 works very well and correct the band gap i.e. in agreement with the previous reports (see Table \ref{tbl:3}). This shows the reliability of hybrid functional HSE06 to predict the electronic structure accurately. For optical properties, we go beyond the HSE06, and performed MBPT viz. GW and BSE calculations. GW calculations determine the fundamental band gap analogous to photoelectron spectroscopy (PES) and inverse photoelectron spectroscopy (IPES)~\cite{PhysRev.139.A796,PhysRevLett.55.1418}, and assumed to be more accurate where as BSE calculation predict the optical band gap akin to experimental optical absorption spectroscopy~\cite{PhysRevLett.80.4510,PhysRevLett.81.2312}. 

\begin{figure}[h]
	\centering
	\includegraphics[width=1.0\textwidth]{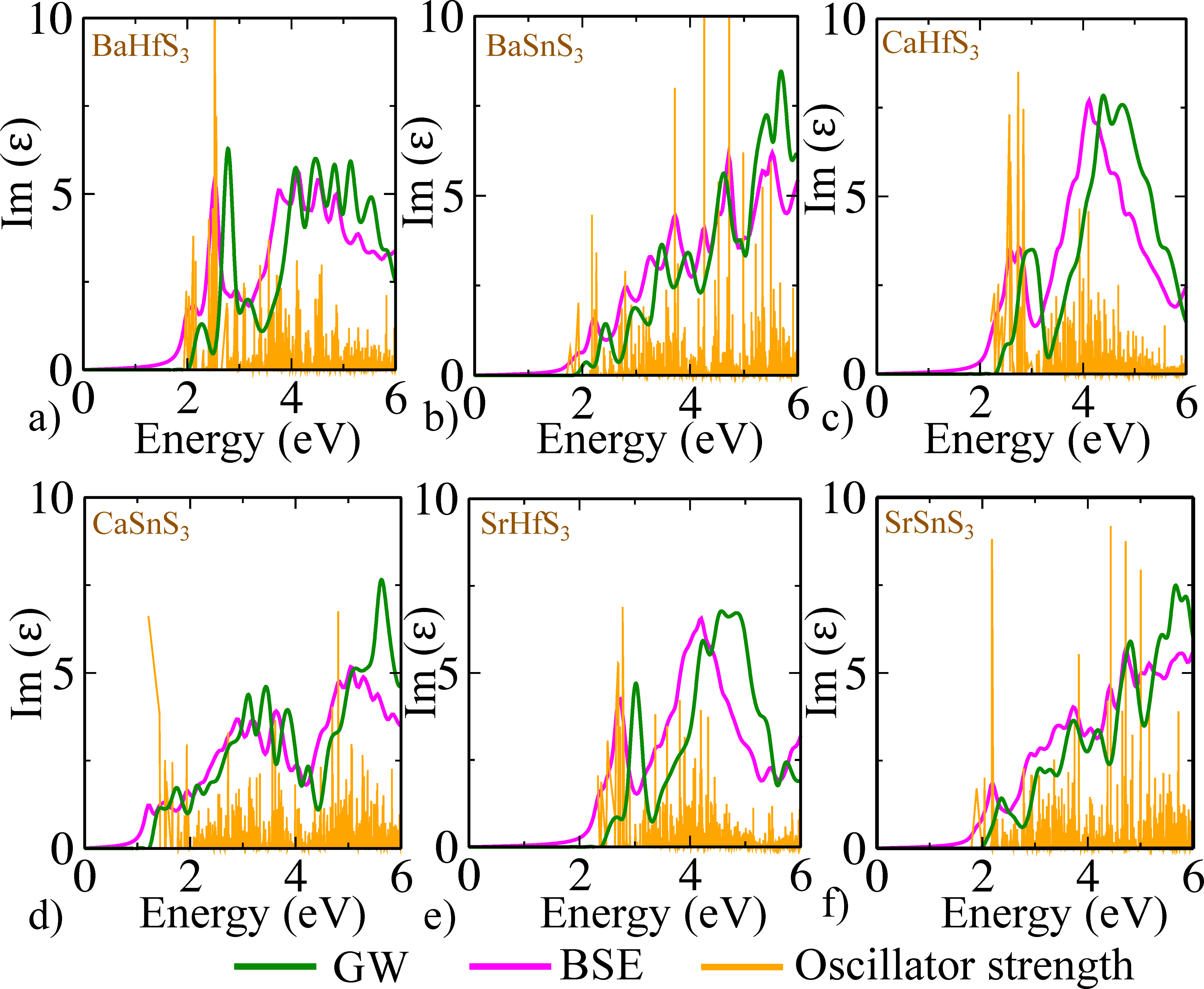}
	\caption{Imaginary [Im $(\upvarepsilon)$] part of the dielectric function average along x, y and z directions for (a) BaHfS$_3$, (b) BaSnS$_3$, (c) CaHfS$_3$, (d) CaSnS$_3$, (e) SrHfS$_3$, (f) SrSnS$_3$ obtained using G$_0$W$_0$@PBE and BSE@G$_0$W$_0$@PBE. The oscillator strength is represented by orange color.}
	\label{fig_optical}
\end{figure}

The electron-hole interaction is explicitly considered in the BSE calculations \cite{doi:10.1021/acs.jpcc.7b07473}. Firstly, we perform single shot GW (G$_0$W$_0$) on the top of PBE to determine the optical response. Note that the quasiparticle (QP) band gap is slighly overestimated as compare to experimental band gap, due to non-inclusion of electron-hole interaction. Table \ref{tbl:3} presents the QP gaps of ABS$_3$ calculated using GW@PBE. The experimental band gaps are nicely captured from GW calculations, however overestimated from the reported theoretical band gap calculations, as many of them are performed using semi-local or hybrid $\epsilon_\textrm{xc}$ functionals. This indicates the advantage of performing MBPT calculations and could be very useful for ongoing experimental work. The imaginary part of the dielectric function [Im ($\upvarepsilon$)] calculated using GW@PBE and BSE@GW@PBE is shown in Figure \ref{fig_optical}. The oscillator strength corresponding to BSE@GW@PBE is also shown in Figure \ref{fig_optical}. We compute the exciton binding energy (E$_\textrm{B}$) by taking the difference of the QP band gap (GW@PBE peak position) and the optical band gap (BSE@GW@PBE peak position). The E$_\textrm{B}$ of the first bright exciton for BaHfS$_3$, BaSnS$_3$, CaHfS$_3$, CaSnS$_3$, SrHfS$_3$ and SrSnS$_3$ is 0.193 eV, 0.134 eV, 0.250 eV, 0.222 eV, 0.256 eV, and 0.148 eV, respectively. From the analysis of BSE eigenvalues, we observe a dark exciton (optically inactive) below the first bright exciton for the case of BaHfS$_3$, BaSnS$_3$, and SrSnS$_3$ perovskites. The E$_\textrm{B}$ corresponding to the dark excitons are 0.236 eV, 0.316 eV, and 0.248 eV for BaHfS$_3$, BaSnS$_3$, and SrSnS$_3$, respectively. The oscillator strength matches well with the excitonic peak position. The optical transitions below the quasiparticle band gap may lead to various interesting excitonic properties\cite{basera2021capturing,jain2021theoretical}. Using aforementioned quantities such as exciton binding energy, band gap, dielectric function, and reduced mass, we compute several excitonic parameters such as excitonic temperature (T$_{exc}$), radius (r$_{exc}$), and probability of wavefunction ($|\phi_\textrm{n}(0)|^2$) for electron-hole pair at zero separation (see details in SI). The inverse of $|\phi_\textrm{n}(0)|^2$) gives a qualitative description of the excitonic lifetime ($\tau$). 
Therefore, the exciton lifetime ($\tau$) for the perovskites follows an order: SrSnS$_3$ $>$ BaSnS$_3$ $>$ CaSnS$_3$ $>$ BaHfS$_3$ $>$ CaHfS$_3$ $>$ SrHfS$_3$. Hence, we obtain that Sn based chalcogenides have larger exciton lifetime than Hf based chalcogenides. Among Sn-based chalcogenides, needle-like phase show large exciton lifetime than orthorhombic phase.
\begin{table}
	\caption{Calculated excitonic parameters for chalcogenide perovskites} 
	\begin{center}
		\begin{tabular}[c]{ccccccc} \hline
			Excitonic parameters  & BaHfS$_3$ &  BaSnS$_3$ & CaHfS$_3$ & CaSnS$_3$ & SrHfS$_3$ & SrSnS$_3$ \\ \hline
			E$_\textrm{B}$ (eV)  & 0.193 & 0.134 & 0.250 & 0.222 & 0.256 & 0.148  \\ 
			T$_\textrm{exc}$ (K)  & 2239 & 1555 & 2901& 2576&2970 & 1717 \\ 
			r$_\textrm{exc}$ (nm) & 0.932 & 1.43  & 0.841 & 1.046& 0.788 & 1.667  \\ 
			$|\phi_\textrm{n}(0)|^2 (10^{27}$m$^{-3})$  & 0.393 & 0.108 & 0.535 & 0.277 & 0.649 & 0.069  \\
			\hline
		\end{tabular}
		\label{tbl:4}
	\end{center}
\end{table}

\begin{figure}[h]
	\centering
	\includegraphics[width=1.0\textwidth]{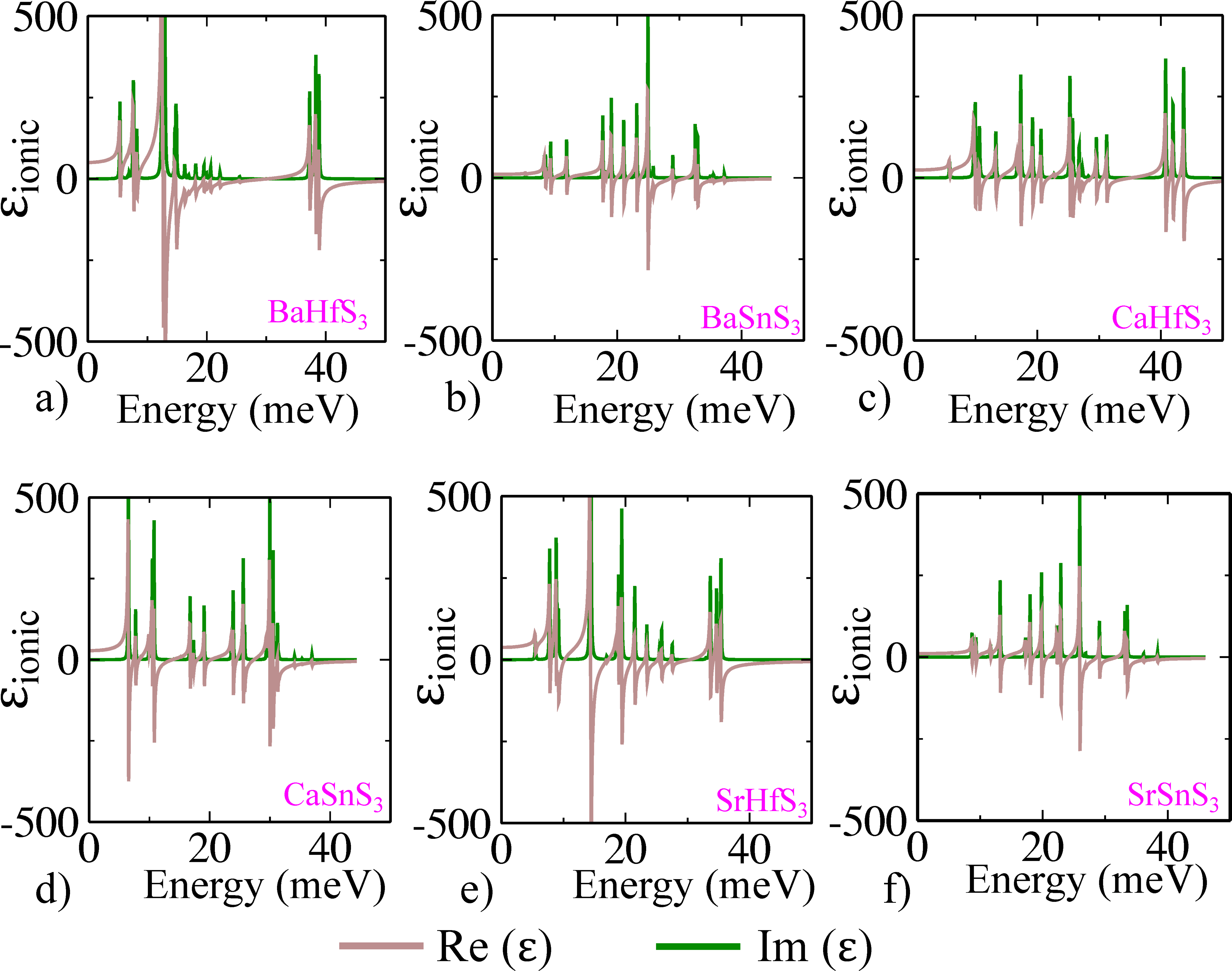}
	\caption{DFPT calculations to determine the ionic contribution to dielectric function for (a) BaHfS$_3$, (b) BaSnS$_3$, (c) CaHfS$_3$, (d) CaSnS$_3$, (e) SrHfS$_3$, (f) SrSnS$_3$.}
	\label{fig_ionic}
\end{figure}

We have obtained high E$_\textrm{B}$ values compared to conventional 3D halide perovskite\cite{doi:10.1021/acs.jpclett.8b02811}. This can be explained by the interplay of electronic and ionic contribution to the dielectric screening. Note that, an electronic contribution is dominant over the ionic one, if E$_\textrm{B}$ is much larger than the longitudinal optical phonon energy ($\hbar\omega_\textrm{LO}$)~\cite{bokdam2016role}. In that scenario, ionic contribution can be neglected, and as a result, E$_\textrm{B}$ does not change. Following this, in chalcogenide perovskite cases, all the dominant longitudinal optical active phonon modes are below 50 meV (see Figure \ref{fig_ionic}). Table \ref{tbl:4} presents E$_\textrm{B}$ calculated from BSE calculations, where E$_\textrm{B} \gg \hbar\omega_\textrm{LO}$, hence ionic screening to dielectric function can be ignored. Further, we have also employed Wannier-Mott approach to validate the E$_\textrm{B}$. As per this model, the E$_\textrm{B}$ is dependent on the reduced mass of the charge carriers ($\mu$) for screened Coulomb interacting e-h pairs and the effective dielectric constant ($\varepsilon_{eff}$). The expression is given by:
\begin{equation}
\textrm{E}_\textrm{B}=\frac{\mu}{\varepsilon_{eff}^2}\textrm{R}_\infty
\label{eq}
\end{equation}
\begin{table}
	\caption{The upper and lower limit of exciton binding energy (E$_\textrm{B}$) for chalcogenide perovskites}
	\label{tbl:bound}
	\centering
	\setlength\extrarowheight{+4pt}
	\begin{tabular}[c]{ccc}
		\hline
		Configurations & Upper bound (eV)& Lower bound (meV)\\
		\hline
		BaHfS$_3$ & 0.188 & 1.24  \\
		BaSnS$_3$ & 0.149 & 19.92  \\
		CaHfS$_3$ &0.227 & 5.47 \\
		CaSnS$_3$ & 0.189 & 3.61 \\
		SrHfS$_3$ & 0.230 & 2.37 \\
		SrSnS$_3$ & 0.138 & 17.27 \\
		\hline
	\end{tabular}
\end{table} 
where, $\textrm{R}_\infty$ and $\varepsilon_{eff}$ are the Rydberg constant and the value lies in between the static value of electronic and ionic dielectric constants, respectively. The upper and lower bound to the exciton binding energy can be calculated from the contribution of electronic and ionic static dielectric constants. For BaHfS$_3$, BaSnS$_3$, CaHfS$_3$, CaSnS$_3$, SrHfS$_3$ and SrSnS$_3$, the corresponding static electronic dielectric constants are 4.10, 4.60, 3.73, 4.09, 3.71, and 4.79, respectively. Note that, these values are obtained using BSE. The DFPT calculations are performed to determine static ionic dielectric constants i.e. 50.45, 10.77, 24.16, 27.91, 37.76, and 10.94 for respective perovskites (see Figure \ref{fig_ionic}). We notice that the needle-like phase has relatively a low value of static ionic dielectric constants. Using Equation~\ref{eq}, we determine the upper and lower bounds of E$_\textrm{B}$ by substituting reduced mass from Table \ref{tbl:2}, and the above mentioned static electronic and ionic dielectric constants. We have found that the upper bounds E$_\textrm{B}$ (Table \ref{tbl:bound}) agree well with the difference of GW and BSE peak positions (listed in Table \ref{tbl:4}). This  confirms that for chalcogenide perovskites, the electronic contribution is dominant over the ionic contribution in dielectric screening.\\
Further, we have used Fr\"{o}hlich mesoscopic model~\cite{C6MH00275G,PhysRevB.96.195202} to investigate the interaction of longitudinal optical phonon modes with the carriers that strongly influence the carrier mobility. This interaction is defined by the dimensionless Fr\"{o}hlich parameter $\alpha$, given by
\begin{equation}
\alpha=\left(\frac{1}{\upvarepsilon_\infty}-\frac{1}{\upvarepsilon_\textrm{static}}\right)\sqrt{\frac{\textrm{R}_\infty}{ch\omega_\textrm{LO}}}\sqrt{\frac{\textrm{m}^*}{\textrm{m}_\textrm{e}}}
\label{eq2}
\end{equation}
where $h$ and $c$ are Planck's constant and speed of light, respectively. The Fr\"{o}hlich parameter $\alpha$ is dependent on the $\upvarepsilon_\infty$ (electronic dielectric constant), $\upvarepsilon_\textrm{static}$ (ionic static dielectric constant), carrier effective mass (m$^*$), and phonon frequency $\omega_\textrm{LO}$. The characteristic frequency $\omega_\textrm{LO}$ has been calculated using athermal `B' scheme of Hellwarth \textit{et al}~\cite{PhysRevB.60.299}, where spectral average of infrared active optical phonon modes are considered. The computed $\alpha$ values are listed in Table~\ref{tbl:ep}, the $\alpha$ parameter for hole is given in SI.
By knowing $\alpha$, we compute the polaron energy (E$_p$) using Equation~\ref{eqp}. Note that the polaron formation may lead to reduction of quasiparticle energy of electron and hole, and is given by the following equation~\cite{bokdam2016role,franchini2021polarons}:
\begin{equation}
\textrm{E}_\textrm{p} = (-\alpha-0.0123\alpha^2)\hbar\omega_\textrm{LO}
\label{eqp}
\end{equation}
For BaHfS$_3$, BaSnS$_3$, CaHfS$_3$, CaSnS$_3$, SrHfS$_3$, SrSnS$_3$, the QP gap is lowered by 0.17, 0.09, 0.21, 0.15, 0.19, 0.08 eV, respectively. By comparing these values with E$_\textrm{B}$, we obtain that the charge separated polaronic state is less stable than the bound exciton. 
Other polaron parameters such as effective polaron masses ($\textrm{m}_\textrm{P}$), radii (l$_\textrm{p}$) and polaron mobilities ($\mu_{\textrm{P}}$) are determined using Hellwarth polaron model~\cite{PhysRevB.60.299}. These parameters determine an upper limit for charge carrier mobilities by assuming only the involvement of optical phonons (see details in SI). However, one may expect a lower value of polaron carrier mobility, if local distortions due to polarons and defect scattering due to acoustic phonons are taken into account. From Table~\ref{tbl:ep}, we have found that large polarons (l$_\textrm{p}$) are formed, and Sn-based chalcogenides possess a large value of polaron mobility than Hf based ones. 
\begin{table}
	\caption{Polaron parameters for electrons in chalcogenide perovskites}
	\label{tbl:ep}
	\centering
	\setlength\extrarowheight{+4pt}
	\begin{tabular}[c]{ccccccc}
		\hline
		Configurations & 1/$\epsilon^*$ &$\omega_\textrm{LO}$ (cm$^{-1}$)&$\alpha$ $_e$&$\textrm{m}_\textrm{P}$ &  l$_{\textrm{P}}$ (\AA) & $\mu_{\textrm{P}}$ (cm$^2$V$^{-1}$s$^{-1}$) \\
		\hline
		BaHfS$_3$ &0.223& 148.53 & 3.68 & 0.76 & 35.75 & 15.82 \\
		BaSnS$_3$ &0.124&195.32  & 2.36& 1.05 &55.26&16.41\\
		CaHfS$_3$ &0.226& 253.92 & 3.13 &0.976&57.96&13.84 \\
		CaSnS$_3$ &0.208& 167.54 & 3.31 &0.766 &41.36&17.01\\
		SrHfS$_3$ &0.242& 158.47 & 4.26& 1.10&34.71&9.58\\
		SrSnS$_3$ &0.117& 199.23 & 1.99 &0.80& 61.06 & 25.08\\
		\hline
	\end{tabular}
\end{table}


\begin{figure}[h]
	\centering
	\includegraphics[width=0.9\textwidth]{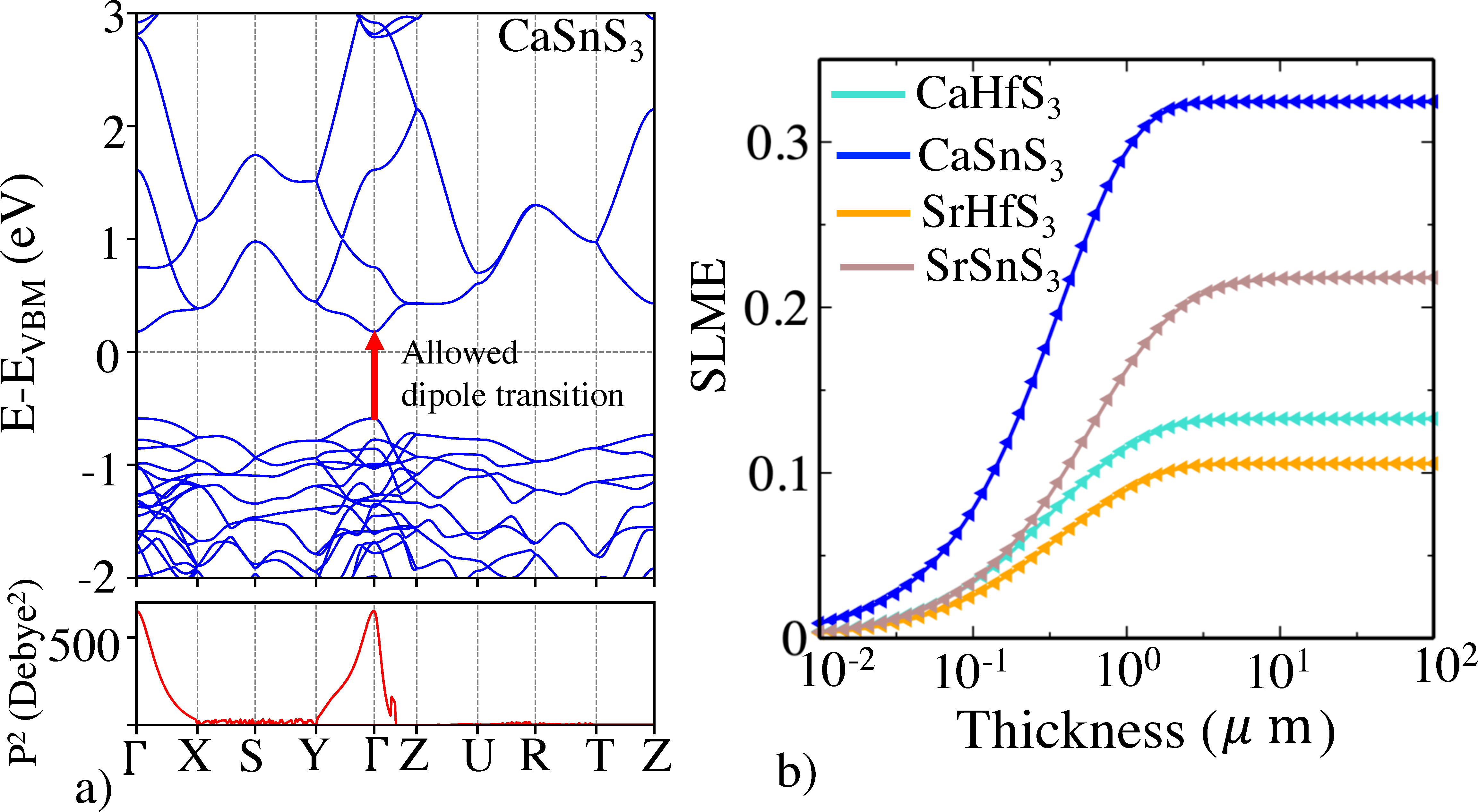}
	\caption{Electronic band structure and transition probability from VBM to CBm (square of dipole transition matrix elements) for CaSnS$_3$. Spectroscopic limited maximum efficiency of ABS$_3$ (A=Ca, Sr, and B=Hf, Sn).}
	\label{fig_slme}
\end{figure}
Notably, ABS$_3$ perovskites can be considered as good solar absorbers as they exhibit large absorption coefficient and most of them show a direct band gap in visible region range. Therefore, to quantify this, we choose a parameter introduced by Yu \textit{et al.} known as spectroscopic limited maximum efficiency (SLME) ~\cite{PhysRevLett.108.068701,slme}. SLME determines the maximum limit of the solar power conversion efficiency of an absorber material. It incorporates the shape of absorption spectra,  band gap and its nature (direct or indirect), thickness of the thin film absorber layer and the material dependent non-radiative recombination losses. This is considered as an improved version of Shockley and Queisser (SQ) efficiency~\cite{doi:10.1063/1.1736034}.  
From above discussion we know that SLME depends on the nature of the band gap, however for various perovskites~\cite{kangsabanik2018double}, it has been seen that even after having a direct electronic band gap, the optical allowed dipole transition from VBM to CBm was forbidden. This happens because of the presence of inversion symmetry that leads to VBM and CBm at the same parity. We, therefore, check the possibility of the optical transition from VBM to CBm for our systems. For this, we calculate the dipole transition matrix, defined as the electric dipole moment related with a transition between the initial state and the final state.  Its square gives the probability of transition between the two states. We have computed the square of the dipole transition matrix elements (p$^2$) for all the systems. We have found that BaHfS$_3$ has forbidden dipole transition at $\Gamma$, regardless of direct electronic band gap (see SI, Fig S1). Figure \ref{fig_slme}, shows an optically allowed dipole transition for CaSnS$_3$ perovskites (for other configurations, see SI, Fig S1). Thus, keeping this in mind, we proceed next for SLME calculations, only for systems, that posseses direct band gap along with optically allowed dipole transitions. Note that, SrSnS$_3$ has also been considered, as the difference of indirect and direct band gap of only few meV's, that does not impact the absorption coefficient. For SLME, the input parameters are band gap, thickness, absorption coefficient and the standard solar spectrum. Figure \ref{fig_slme} presents the calculated SLME of ABS$_3$ (A=Ca, Sr, and B=Hf, Sn). SLME values at 10 $\mu$m absorber layer thickness are 13.26\%, 32.45\%, 10.56\%, and 21.80\% for CaHfS$_3$, CaSnS$_3$, SrHfS$_3$, SrSnS$_3$ respectively (see Figure \ref{fig_slme}). We have achieved the highest value of SLME for CaSnS$_3$, that could serve as a best material for photovoltaic applications. 

In summary, we have determined the electronic, optical (excitonic and polaronic) properties of chalcogenide perovskites ABS$_3$ (A= Ba, Ca, Sr, and B=Hf, Sn) under the framework of DFT and MBPT. We have obtained reduce mass, $\mu$ $<$ 1 for all the cases, that confirms high carrier mobility. For BaSnS$_3$ and SrSnS$_3$, the difference in Sn-S bonding strength of SnS$_6$ octahedra leads to lighter valence bands, which in turn boost up the hole mobility. The large value of exciton binding energy is noted for chalcogenide perovskites i.e., 0.193, 0.134,0.250, 0.222, 0.256, and 0.148 eV for BaHfS$_3$, BaSnS$_3$, CaHfS$_3$, CaSnS$_3$, SrHfS$_3$ and SrSnS$_3$, respectively. The exciton binding energy computed from BSE calculations are well in agreement with Wannier-Mott approach, thereby indicate that electronic contribution is dominant over ionic contribution in dielectric screening. Further, from Fröhlich’s mesosocopic model, we have found that the charge-separated polaronic states are less stable than the bound excitons. Lastly, the spectroscopic limited maximum efficiency (SLME) suggest that among all studied chalcogenide perovskites, CaSnS$_3$ could be a best choice for efficient photovoltaic applications. 

\section{Computational Methods}
The DFT~\cite{PhysRev.136.B864,PhysRev.140.A1133} calculations are performed as implemented in the Vienna \textit{ab initio} simulation package (VASP)~\cite{KRESSE199615,PhysRevB.59.1758}. Projector-augmented wave (PAW) pseudopotentials~\cite{PhysRevB.50.17953,PhysRevB.59.1758} are used to describe the ion-electron interactions. The valence states considered for Ca, Sr, Ba, Hf, Sn and S are 3s$^2$3p$^6$4s$^2$,  4s$^2$4p$^6$5s$^2$, 5s$^2$5p$^6$6s$^2$, 5p$^6$6s$^2$6d$^4$, 4d$^{10}$5s$^2$5p$^2$, and 3s$^2$3p$^4$, respectively. PBE~\cite{PhysRevLett.77.3865} exchange-correlation ($\epsilon_\textrm{xc}$) functional has been used for the structural optimization. The structural relaxation are performed until  the forces are smaller than 0.001 eV/\AA. The plane wave basis set expansion or kinetic energy cutoff is set to 500 eV. For single point energy calculations, the tolerance criteria for energy convergence is set to 0.001 meV. For Brillouin zone integration, a $k$-grid of $7\times7\times5$ is used, generated from Monkhorst-Pack~\cite{PhysRevB.13.5188} scheme. The advanced hybrid $\epsilon_\textrm{xc}$ functional HSE06~\cite{doi:10.1063/1.2404663} and GW calculations are performed for the better estimation of the electronic band gap. To determine excitonic properties, Bethe-Salpeter equation (BSE)~\cite{PhysRevLett.80.4510,PhysRevLett.81.2312} calculations are performed on top of single shot GW~\cite{PhysRev.139.A796,PhysRevLett.55.1418} (G$_0$W$_0$). PBE $\epsilon_\textrm{xc}$ functional is considered as a starting point for single shot GW calculations. A grid of 50 frequency points is used for the polarizability calculations. We have considered sufficient number of unoccupied bands i.e. nine times the number of occupied bands. The convergence for number of unoccupied bands is shown in SI. For BSE calculations, $\Gamma$-centered $3\times3\times2$ $k$-grid has been used. The electron-hole kernel for BSE calculations are constructed by 24 occupied and 24 unoccupied bands (for convergence see SI). Density functional perturbation theory (DFPT) calculations are performed to get the ionic contribution to dielectric screening using $7\times7\times5$ $k$-grid. Note that spin-orbit coupling (SOC) has not been taken into account, as it does not alter the electronic structure (see SI).
\begin{acknowledgement}
PB acknowledges UGC, India, for the senior research fellowship [grant no. 20/12/2015 (ii) EUV].
SB acknowledges the financial support from SERB under core research grant (grant no. CRG/2019/000647). We acknowledge the High Performance Computing (HPC) facility at IIT Delhi for computational resources.
\end{acknowledgement}
\begin{suppinfo}
(I) Electronic band structure and transition probability. (II) Effect of spin-orbit coupling on band gap. (III) Convergence of number of bands used in single shot GW calculations. (IV) Convergence of number of occupied (NO) and unoccpied bands (NV) used in electron-hole interaction kernel for BSE calculations. (V) Excitonic parameters: exciton binding energy (E$_\textrm{B}$), dielectric constant ($\upvarepsilon_\textrm{eff}$), band gap (E$_g$) and reduced mass ($\mu$). (VI) Polaron parameters of chalcogenide perovskites. (VII) pDOS of S atoms of chalcogenide perovskites.
\end{suppinfo}
\bibliography{achemso-demo}
\end{document}


\begin{center}
{\Large \bf Supplemental Material}\\ 
\end{center}
\begin{enumerate}[\bf I.]
	
\item Electronic band structure and transition probability
\item Effect of spin-orbit coupling on band gap
\item Convergence of number of bands used in single shot GW calculations
\item Convergence of number of occupied (NO) and unoccpied bands (NV) used in electron-hole interaction kernel for BSE calculations
\item Excitonic parameters: exciton binding energy (E$_\textrm{B}$), dielectric constant ($\upvarepsilon_\textrm{eff}$), band gap (E$_g$) and reduced mass ($\mu$)
\item Polaron parameters of chalcogenide perovskites
\item pDOS of S atoms of chalcogenide perovskites

\end{enumerate}
\vspace*{12pt}
\clearpage
\newpage

\section{Electronic band structure and transition probability}
\begin{figure}[H]
	\centering
	\includegraphics[width=1.1\textwidth]{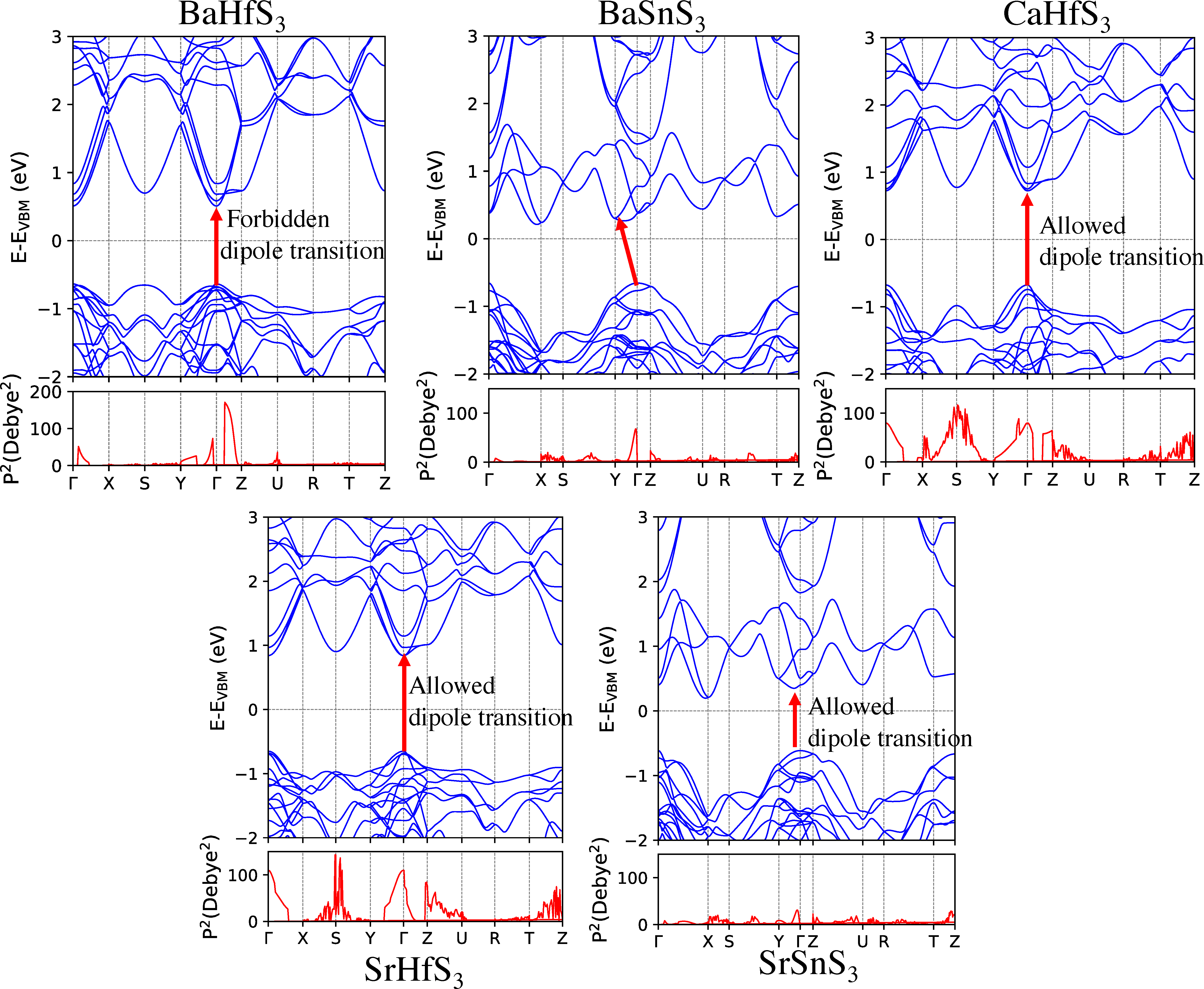}
	\caption{ Electronic band structure and transition probability from VBM to CBm (square of dipole transition matrix elements) using PBE $\epsilon_\textrm{xc}$ functional. BaHfS$_3$ has forbidden dipole transition at $\Gamma$ point.}
	\label{fig1}
\end{figure}
\section{Effect of spin-orbit coupling on band gap}
\begin{table}[H]
	\caption{Band gap of BaHfS$_3$, BaSnS$_3$, CaHfS$_3$, CaSnS$_3$, SrHfS$_3$, and SrSnS$_3$ using PBE $\epsilon_\textrm{xc}$ functional.}
	\label{tbl:bound}
	\centering
	\setlength\extrarowheight{+4pt}
	\begin{tabular}[c]{ccc}
		\hline
		Configurations & Without SOC (eV) & With SOC (eV)\\
		\hline
		BaHfS$_3$ & 1.12 & 1.03  \\
		BaSnS$_3$ & 0.87 (0.95) & 0.86 (0.94)  \\
		CaHfS$_3$ &1.38 & 1.31 \\
		CaSnS$_3$ & 0.76& 0.79 \\
		SrHfS$_3$ & 1.47 & 1.37 \\
		SrSnS$_3$ & 0.96 & 0.95 \\
		\hline
	\end{tabular}
\end{table} 
\section{Convergence of number of bands used in single shot GW calculations}
\begin{table}[H]
	\caption{Band gap (in eV) of BaHfS$_3$ using single shot GW@PBE with different number of bands.}
	\label{tbl:S2}
	\centering
	\setlength\extrarowheight{+4pt}
	\begin{tabular}[c]{ccccccc}
		\hline
		Number of bands & BaHfS$_3$\\
		\hline
		320 & 2.19\\
		480 & 2.16   \\
	    640 &2.17 \\
		720 & 2.17\\
		800 & 2.17 \\
		
		\hline
	\end{tabular}
\end{table} 
Table~\ref{tbl:S2} shows that 640 bands are sufficient for the calculations. For our calculations, we have used 720 number of occupied bands and 500 eV plane wave energy cutoff.

\section{Convergence of number of occupied (NO) and unoccpied bands (NV) used in electron-hole interaction kernel for BSE calculations}
\begin{figure}[H]
	\centering
	\includegraphics[width=0.8\textwidth]{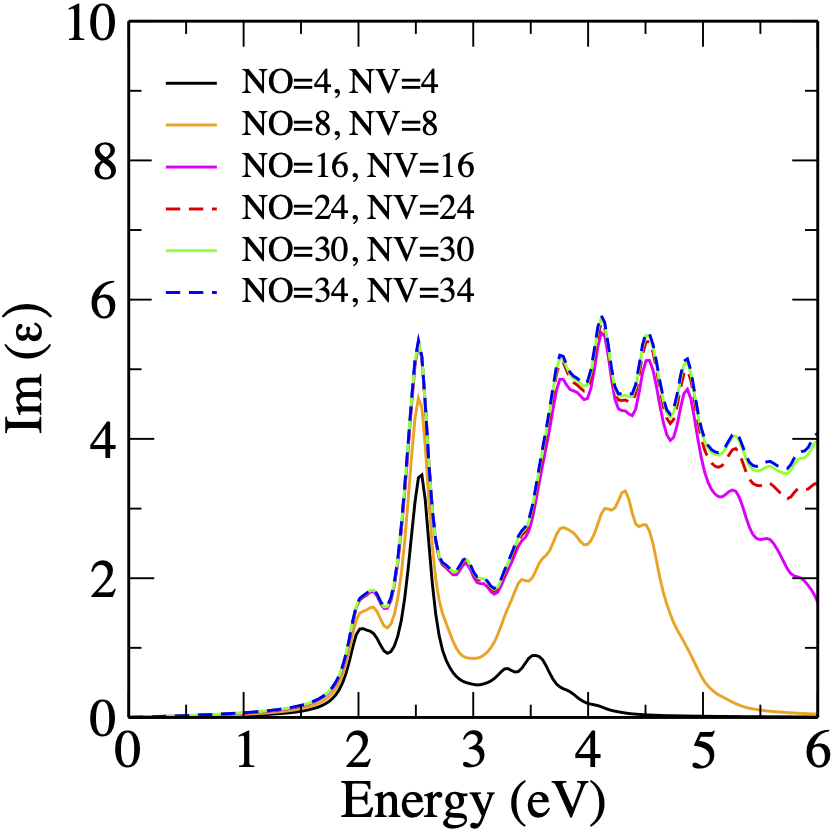}
	\caption{Imaginary part of electronic dielectric function for BaHfS$_3$ with different number of valence (NO) and conduction bands (NV) used in electron-hole interaction kernel.}
	\label{fig1}
\end{figure}
From Fig S2, we observe that the imaginary part of the dielectric function is converged for NO=NV=16 bands. We have used NO=NV=24 bands in our calculations. 

\section{Excitonic parameters: exciton binding energy (E$_\textrm{B}$), dielectric constant ($\upvarepsilon_\textrm{eff}$), band gap (E$_g$) and reduced mass ($\mu$)}
The thermal energy needed to separate the exciton is given by~\cite{basera2021capturing}:
\begin{equation}
\mathrm{T}_{\mathrm{exc}}=\frac{\mathrm{E}_{\mathrm{B}}}{\mathrm{k}_{\mathrm{B}}}
\end{equation}
where E$_\textrm{B}$ and k$_\textrm{B}$ are the exciton binding energy and the Boltzmann constant, respectively.
The exciton radius with the orbital n is given by:
\begin{equation}
r_{\text {exc }}=\frac{m_{0}}{\mu} \varepsilon_{\text {eff }} n^{2} r_{\mathrm{Ry}}
\end{equation}
where, $r_{\mathrm{Ry}} = 0.0529$ nm is the Bohr radius. m$_0$ and $\mu$ are the free electron mass and reduced mass. $\varepsilon_{\text {eff }}$ is defined as an intermediate value between the static electronic and ionic dielectric constant. In our case, we have considered electronic dielectric constant, since electronic contribution is dominant over ionic contribution in the dielectric screening. The inverse of probability of wave function for e-h pair at zero separation ($|\phi_\textrm{n}(0)|^2$) gives a qualitative description of the excitonic lifetime ($\tau$). The $\left|\phi_{n}(0)\right|^{2}$ is given by:

\begin{equation}
\left|\phi_{n}(0)\right|^{2}=\frac{1}{\pi\left(r_{e x c}\right)^{3} n^{3}}
\end{equation}

\section{Polaron parameters of chalcogenide perovskites}
\begin{table}
	\caption{Electron-phonon coupling parameters for holes in chalcogenides perovskites}
	\label{tbl:bound}
	\centering
	\setlength\extrarowheight{+4pt}
	\begin{tabular}[c]{ccc}
		\hline
		Configurations & $\alpha_h$\\
		\hline
		BaHfS$_3$ & 4.87   \\
		BaSnS$_3$ & 1.42 \\
		CaHfS$_3$ & 3.32  \\
		CaSnS$_3$ & 3.57  \\
		SrHfS$_3$ &  4.81  \\
		SrSnS$_3$ & 1.27 \\
		\hline
	\end{tabular}
\end{table} 
 ‘‘Frohlich’s mesoscopic model’’is used to investigate the electron–phonon interaction in our system. In 1954, Frohlich introduced an interaction parameter that describes theoretically the motion of an electron in the vicinity of polar lattice vibration. This parameter is a dimensionless quantity known as Frohlich coupling constant~\cite{frohlich1954electrons,feynman1962mobility}.
\begin{equation}
\begin{split}
\alpha = \frac{1}{\epsilon^*}\sqrt{\frac{\textrm{R}_\textrm{y}}{{ch}\omega_{\textrm{LO}}}}\sqrt{\frac{\textrm{m}^*}{\textrm{m}_\textrm{e}}}
\label{eq3}
\end{split}
\end{equation}
where coupling constant $\alpha$ quantifies the strength of electron-phonon coupling. m$^*$, m$_\textrm{e}$, $h$ and $c$  are the effective mass of electron, rest mass of the electron, Planck's constant, and speed of light, repectively. $\omega_{\textrm{LO}}$ (in [cm$^{-1}$] units) is the optical phonon frequency, $1/\epsilon^*$ is the ionic screening parameter ($1/\epsilon^*$ = $1/\epsilon_{\infty}$ - $1/\epsilon_{\textrm{static}}$ were, $\epsilon_{\textrm{static}}$ and $\epsilon_{\infty}$ are static and high frequency dielectric constant) and $\textrm{R}_\textrm{y}$ is the Rydberg energy.  \\
Further, Feynman extended the Frohlich's polaron theory so that the effective mass of polaron\cite{feynman1955slow} (m$_{\textrm{P}}$) can be calculated using the given equation~\cite{feynman1962mobility}.
\begin{equation}
\begin{split}
 \textrm{m}_{\textrm{P}} = \textrm{m}^* \left(1 + \frac{\alpha}{6} +  \frac{\alpha^2}{40} + ......\right)
\label{eq4}
\end{split}
\end{equation}
where m$^*$ is the effective mass computed from GW band structure calculations. This expression shows that polaron mass is always higher than the effective mass calculated from band structure calculations, thereby influencing charge carrier mobilities. \\
Further, we have determined the polaron radii \cite{sendner2016optical} using the following equation:
\begin{equation}
\begin{split}
\textrm{l}_\textrm{P} = \sqrt{\frac{h}{2c\textrm{m}^*\omega_{\textrm{LO}}}}
\label{eq5}
\end{split}
\end{equation} 
Note that, these polaron parameters are used to estimate an upper limit for charge carrier mobilities $\mu$, assuming that only optical phonons are involved. \\
Hellwarth polaron model\cite{hellwarth1999mobility} is used to define the polaron mobility~\cite{frost2017calculating}:
\begin{equation}
\begin{split}
\mu_{\textrm{P}} = \frac{(3\sqrt{\pi}e)}{2\pi {c}\omega_{\textrm{LO}}\textrm{m}^*\alpha} \frac{\textrm{sinh}{(\beta/2)}}{\beta^{5/2}}\frac{w^3}{v^3}\frac{1}{K}
\label{eq5}
\end{split}
\end{equation}
where, $\beta$ = $hc$ $\omega_{\textrm{LO}}$/$k_{\textrm{B}}$T, $e$ is the electronic charge, m$^*$ is the effective mass of charge carrier, $\textit{w}$ and $\textit{v}$ correspond to temperature dependent variational parameters. $\textit{K}$ is a function of $\textit{v}$, $\textit{w}$, and $\beta$\cite{hellwarth1999mobility} i.e., defined as follows:
\begin{equation}
\begin{split}
K(a,b) = \int_0^\infty du[u^2 + a^2 - b\textrm{cos}(vu)]^{-3/2}\textrm{cos}(u)
\label{eq6}
\end{split}
\end{equation}
Here, a$^2$ and b are calculated as:
\begin{equation}
\begin{split}
a^2 = (\beta /2)^2 + \frac{(v^2 - w^2)}{w^2v} \beta coth(\beta v/2)
\label{eq7}
\end{split}
\end{equation} 
\begin{equation}
\begin{split}
b =  \frac{(v^2 - w^2)}{w^2v}\frac{\beta}{\textrm{sinh}(\beta v/2)}
\label{eq8}
\end{split}
\end{equation} 

From the following expressions, we have determined the polaron parameters for chalcogenides perovskites.

\section{pDOS of S atoms of chalcogenide perovskites}
\begin{figure}[H]
	\centering
	\includegraphics[width=1.1\textwidth]{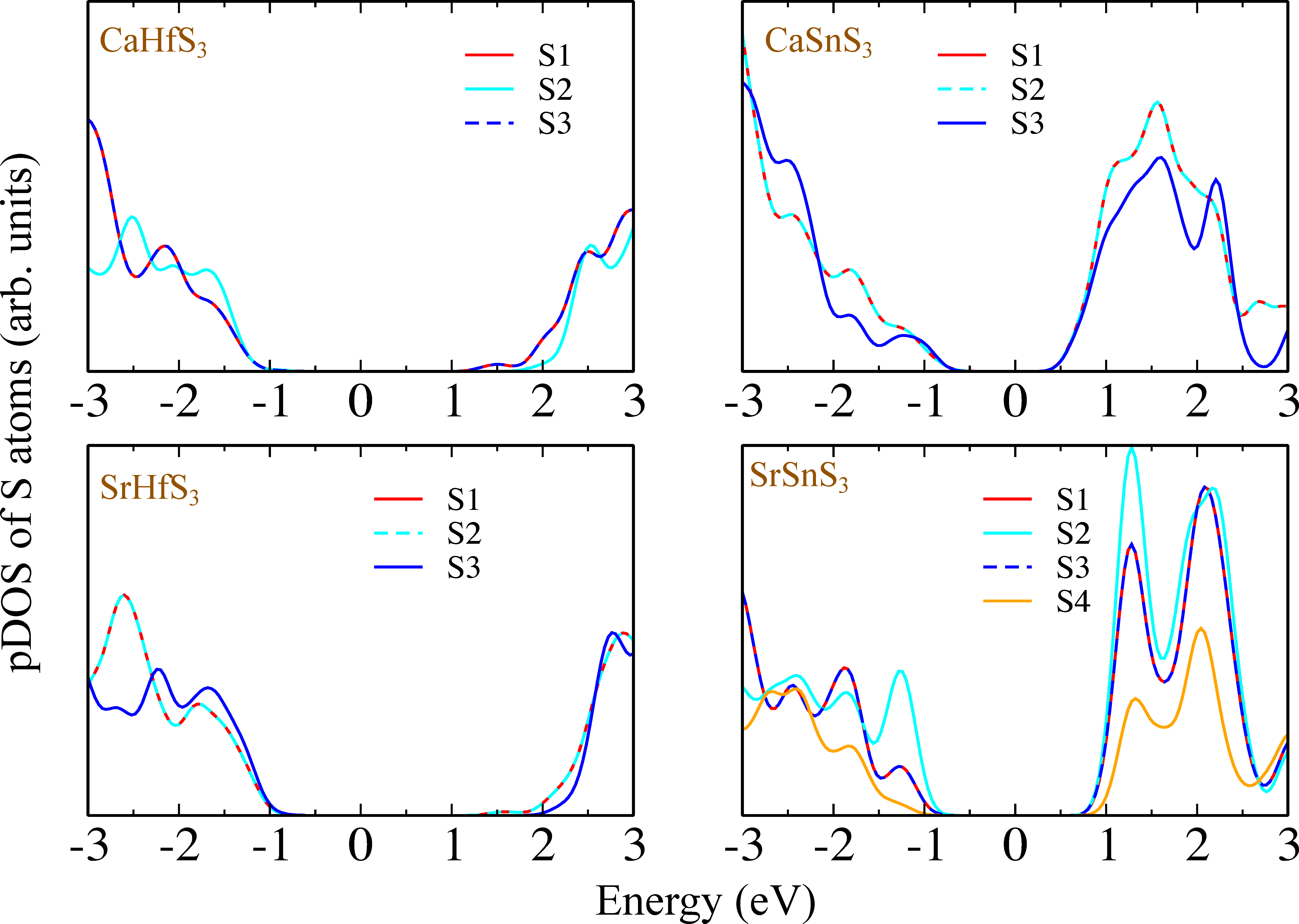}
	\caption{Sulphur atom projected density of states for CaHfS$_3$, CaSnS$_3$, SrHfS$_3$, and SrSnS$_3$.}
	\label{fig1}
\end{figure}

\bibliography{achemsodemo1.bib}